\documentclass[prd,superscriptaddress,nofootinbib,10pt]{revtex4-2}   
\usepackage{epsfig}    
\usepackage{amsmath} 
\usepackage{amsfonts}   
\usepackage{float}    
\usepackage{amssymb}    
\usepackage{slashed} 
\usepackage{color}       
\usepackage{mathtools}
\usepackage{pbox}
\usepackage{subfigure}
\usepackage{array,multirow,makecell}
\usepackage[colorlinks,citecolor=blue, linkcolor = blue, urlcolor = blue]{hyperref}
\usepackage{tabularx}
\usepackage{titlesec} 
\usepackage{scrextend}
\usepackage{ulem}
\usepackage{natbib} 
\usepackage{array,multirow,makecell}
\usepackage{lipsum}

\def\lsim{\raise0.3ex\hbox{$\;<$\kern-0.75em\raise-1.1ex\hbox{$\sim\;$}}}
\def\gsim{\raise0.3ex\hbox{$\;>$\kern-0.75em\raise-1.1ex\hbox{$\sim\;$}}}

\newcommand{\tg}{T_\gamma}
\newcommand{\tn}{T_\nu}
\newcommand{\be}{\begin{equation}}
\newcommand{\ee}{\end{equation}}
\newcommand{\bea}{\begin{eqnarray}}
\newcommand{\eea}{\end{eqnarray}}
\newcommand{\nn}{\nonumber}
\newcommand{\lra}{\leftrightarrow}

\begin{document}

\title{Imprints of MeV Scale Hidden Dark Sector at Planck Data}

\author{Sougata Ganguly}
\email{tpsg4@iacs.res.in}
\affiliation{School of Physical Sciences, Indian Association for the 
Cultivation of Science,\\ 2A $\&$ 2B Raja S.C. Mullick Road, 
Kolkata 700032, India}

\author{Sourov Roy }
\email{tpsr@iacs.res.in}
\affiliation{School of Physical Sciences, Indian Association for the 
Cultivation of Science,\\ 2A $\&$ 2B Raja S.C. Mullick Road, 
Kolkata 700032, India} 

\author{Abhijit Kumar Saha}
\email{psaks2484@iacs.res.in}
\affiliation{School of Physical Sciences, Indian Association for the 
Cultivation of Science,\\ 2A $\&$ 2B Raja S.C. Mullick Road, 
Kolkata 700032, India}

\begin{abstract}
New light species can contribute to
the number of effective relativistic degrees of freedom ($N_{\rm eff}$) at Cosmic Microwave Background (CMB) which is precisely measured by 
Planck. In this work, we consider an MeV scale thermally decoupled non-minimal dark sector
and study the imprint of the dark sector dynamics on the measurement of $N_{\rm eff}$
at the time of CMB formation. We have predicted the allowed region of 
model parameter space in the light of constraints arising from the measurements 
of both $N_{\rm eff}$ and dark matter relic density by Planck. It turns out
that the impact of the dark sector dynamics on $N_{\rm eff} $ is significant  
in case of a non-hierarchical mass spectrum of the dark sector particles.
\end{abstract}

\maketitle

\flushbottom

\section{Introduction} 
\label{sec:intro}
The increasing tension due to non-detection of dark matter at several direct
and indirect detection experiments \cite{XENON100:2012itz,CRESST:2015txj,SuperCDMS:2015eex,Klasen:2015umae,PandaX-II:2016vec} 
motivate to explore alternative scenarios beyond
weakly interacting massive particle (WIMP) paradigm \cite{Feng:2010gw,Roszkowski:2017nbc,
Schumann:2019eaa,Lin:2019uvt,Leane:2020liq}. One of the attractive proposals is freeze in 
mechanism \cite{Hall:2009bx,Bernal:2017kxu} where the DM
abundance results from decays or annihilations of visible sector particles. In this
scenario, the DM never reaches thermal equilibrium with the Standard Model (SM) bath
due to its feeble interaction strength. Such DM candidates are popularly dubbed as feebly 
interacting massive particle (FIMP). Interestingly, the final abundance of 
the FIMP dark matter is sensitive to the initial production 
mechanism in contrast to WIMP. A different kind of scenario has been discussed in 
\cite{Pospelov:2007mp,Feng:2008mu, Cheung:2010gj,Chu:2011be} where the dark matter is a part of a secluded
dark sector and the annihilation of the dark matter into the other dark sector particles
set the relic abundance. This is known as the secluded dark sector
freeze-out.

The presence of any beyond Standard Model (BSM)
light species can leave detectable imprints at Planck satellite experiment. The Planck data provides 
the number of relativistic degrees of freedom
at the time of Cosmic Microwave Background (CMB) to be $N_{\rm eff}^{\rm CMB} =2.99^{+ 0.34}_{-0.33}$
with 95$\%$ confidence limit (C.L) \cite{Planck:2018vyg}. 
In SM alone, an accurate analysis of neutrino decoupling 
by including exact collision terms, effect of neutrino oscillations, and finite temperature 
QED estimates $N_{\rm eff}^{\rm SM}=3.04$ at the time of (CMB) 
\cite{Mangano:2005cc,deSalas:2016ztq, Akita:2020szl, Bennett:2020zkv}.
Earlier efforts of constraining light BSM degrees 
of freedom can be found in \cite{Boehm:2012gr,Brust:2013ova,Nollett:2014lwa,
Wilkinson:2016gsy,Abazajian:2019oqj,DEramo:2018vss,Luo:2020fdt,Luo:2020sho, Biswas:2021kio}. 
These studies are based on the assumptions that SM neutrino decouples instantaneously 
and the new BSM particle is in thermal equilibrium either with the neutrino bath
or with the photon bath. 
A more detailed analysis in \cite{Escudero:2018mvt,Escudero:2019gzq,
Ibe:2019gpv,EscuderoAbenza:2020cmq} by incorporating non-instantaneous SM neutrino decoupling imposes lower bound on masses 
of BSM degrees of freedom from the measurement of 
$N_{\rm eff}^{\rm CMB}$. In addition to that, during Big Bang Nucleosynthesis (BBN), non-negligible abundances of the 
additional light species increase the Hubble parameter 
which in turn may alter the abundances of Helium-4 and Deuterium. The 
measured abundances of Helium-4 and Deuterium give rise to an upper bound
on number of relativistic degrees of freedom at the
time of BBN ($N_{\rm eff}^{\rm BBN}$) to be $2.878\pm 0.278$
with $68.3\%$ C.L \cite{Fields:2019pfx}. 
Thus light species with mass $\lesssim 
\mathcal{O}(1)$MeV can also contribute to the  $N_{\rm eff}^{\rm BBN}$ \cite{Nollett:2014lwa,Depta:2019lbe,Ghosh:2020vti,Giovanetti:2021izc}.

The measurement of effective relativistic degrees of freedom can
be used as a probe of light dark matter as well. A concise discussion on other possible modes of probing a light dark 
sector can be found in \cite{Knapen:2017xzo,Choudhury:2019tss,Choudhury:2019sxt}. 
In \cite{Boehm:2013jpa,Kitabayashi:2015oda,Green:2017ybv,Berlin:2017ftj,Escudero:2018mvt,Berlin:2018ztp,Sabti:2019mhn}, it 
has been argued that a light dark matter can contribute sizeably to $N_{\rm eff}^{\rm CMB}$ if it
remains in thermal contact with either neutrino or electron bath  
till late time. This poses a lower bound on the MeV scale DM which differs 
depending on the spin of the dark matter candidate.  

In this work, we consider an MeV scale non-minimal dark sector consisting of two real gauge
singlet scalar fields. The lightest scalar is assumed to be neutrinophilic. The heavier
scalar, being stable in the Universe lifetime can be identified as a viable dark matter
candidate. Both the scalars interact very feebly with the visible sector and hence never
attain thermal equilibrium with the SM bath, thus form a secluded dark sector. First, the non-thermal production of lighter
neutrinophilic scalar takes place from the inverse decay of SM neutrinos and subsequently
its annihilation yields the dark matter. The production rate of DM depends on two factors. One is the abundance of the 
mediator particle and secondly DM coupling strength to the mediator also determines the efficiency of DM yield. After 
the production, the dark matter annihilates back to the lighter scalar and  in fact a stronger coupling leads to formation 
of internal dark sector equilibrium with a different temperature other than that of SM. Finally, when the ratio between dark sector
 interaction strength and the expansion rate of the Universe comes down below unity, the
 dark matter decouples and freezes out. 

We anticipate that such a non-trivial dynamics of MeV scale dark sector can have two 
fold impacts on the observation of Planck measurement of $N_{\rm eff}^{\rm CMB}$. Firstly, if the lighter scalar decays to SM 
neutrinos during or after neutrino decoupling, it raises the neutrino bath temperature 
and in turn increases $N_{\rm eff}^{\rm CMB}$. Secondly, when mass hierarchy between dark 
matter and the lighter scalar is small, it may increase the $N_{\rm eff}^{\rm CMB}$ further. 
In order to address such possibilities, we have performed a detailed study by solving system of coupled differential 
equations for the evolutions of number densities and 
the temperatures for the relevant species. We also include constraints on DM 
relic density in our present analysis. The bound on the measurement
of $\Delta N_{\rm eff}^{\rm CMB}$ are expected to be more stringent with the next generation CMB 
experiments with improved sensitivity. In fact the proposed CMB Stage IV (CMB-S4) 
experiment \cite{Abazajian:2019eic} has the ability to strengthen 
the upper bound of $\Delta N_{\rm eff}^{\rm CMB}$ upto 0.06 at 95\% C.L 
and can potentially probe our proposed setup.

\section{Basic set up}
\label{sec:basic_set_up}
In this section, we describe the particle contents of our proposed set up and their 
interaction pattern. We consider an extension of SM with two real gauge singlet scalar 
fields $S$ and $\phi$. The effective Lagrangian of our interest is given by,
\begin{align}
-\mathcal{L}\supset \frac{1}{2}m_S^2 S^2+ \frac{1}{2} m_\phi^2\phi^2+\frac{\lambda}{4} S^2\phi^2+ 
\dfrac{\kappa}{2} \phi \overline{\nu_i}\nu_i,
\label{eq:lagrangian}
\end{align}
where we have considered the relevant mass scales ($m_S$ and $m_\phi$) and the coupling 
coefficients ($\lambda$ and $\kappa$) as real and positive. In our framework, the scalar field $\phi$ is purely neutrinophilic
whereas $S$ is the DM candidate. The stability of $S$ can be ensured by imposing a $\mathbb{Z}_2$ symmetry under which $S$ has odd charge.
The last term of Eq.\,\ref{eq:lagrangian} describes the interaction of the dark sector with the neutrino bath. Here the couplings 
of $\phi$ with the neutrinos are flavor universal and $i$ represents the SM neutrino flavors. 
In our analysis, we work in the regime $m_\phi<m_S$. 
In the opposite limit where the DM is lighter than the mediator particle, 
the dark matter $S$ has negligible impact on $N_{\rm eff}^{\rm CMB}$. This is because in this case we need smaller values of the portal
coupling to satisfy the relic density constraint, which in turn implies that the impact on $N_{\rm eff}^{\rm CMB}$ is negligible.

The effective vertex $\phi\overline{\nu}\nu$ is clearly not invariant under $SU(2)_L\otimes U(1)_Y$. 
This can be originated from an $SU(2)_L\otimes U(1)_Y$ gauge invariant set up which is valid at higher energy scale. 
In appendix\,\ref{sec:appUV}, we have discussed two such realisations, guided by the introduction of a
discrete symmetry $\mathbb{Z}_4$ in both cases, under which $\phi$ transforms non-trivially. This choice automatically
forbids the interaction terms involving $\phi$ and other SM particles at linear order in $\phi$.
Additionally, there exist other renormalisable terms such as $\phi^2|\Phi|^2,\,S^2|\Phi|^2$ in the Lagrangian,
where $\Phi$ is the SM Higgs doublet. These may lead to new production channels for the dark sector (DS) particles. 
However, in the present work we are interested in a specific scenario 
where the DS gets populated dominantly from the SM neutrinos (for example in \cite{Du:2020avz}). 
This can be realized by either choosing the Higgs portal coupling ($\phi^2 |\Phi|^2$ etc.) parameters to be tiny enough or 
assuming the reheating temperature of the Universe much lower than the Higgs boson mass 
scale \cite{Hannestad:2004px,Hasegawa:2019jsa}, making the Higgs abundance Boltzmann suppressed 
in the early Universe. In both the cases, the DS does not equilibrate with the SM bath through Higgs portal.

\begin{figure*}
\includegraphics[height=3.2cm,width=15cm]{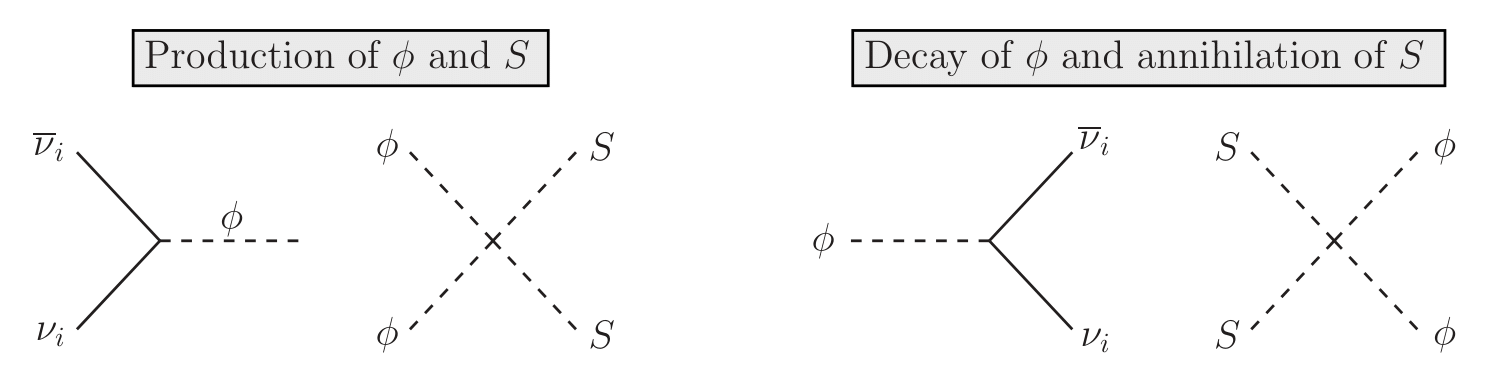} 
\caption{Feynman diagrams indicating productions, annihilations of  
particles as well as decay mode of $\phi$.}
\label{fig:feyn_dia}
\end{figure*}

In the present study, we aim to probe the secluded non-minimal dark sector (DS) with field contents $S$ 
and $\phi$ at Planck experiment. The dark sector is secluded when $S$ and $\phi$ never reach thermal 
equilibrium with the SM bath. The non-thermalisation of dark sector with SM bath implies,
\begin{align}
 \sum_{i} n_{\nu_i}^{\rm eq}(T_{\nu_i})\langle\sigma v \rangle_{\overline{\nu}_i\nu_i\rightarrow \phi}< H,
 \label{eq:nu-phi-eq}
\end{align}
where $\langle\sigma v \rangle_{\overline{\nu}_i\nu_i\rightarrow \phi}$ is the thermally averaged cross section for the 
inverse decay of $\phi$ and $H$ represents the Hubble parameter of the Universe. A conservative check for Eq.\,\ref{eq:nu-phi-eq} 
can be performed at the freeze in temperature $T_{\nu}\sim m_\phi$ of $\phi$ yield. 
Then using the standard analytical expressions for $\langle\sigma v 
\rangle_{\overline{\nu}_i\nu_i\rightarrow \phi}$, $n_{\nu_i}^{\rm eq}$ and Hubble parameter, the Eq.\,\ref{eq:nu-phi-eq} can be converted to,
\begin{align}
\left(\dfrac{m_\phi}{1\rm MeV}\right) > 0.3 \times \left(\dfrac{y}{10^{-10}}\right)^2\,\,,
\label{eq:nu_SM_NT}
\end{align}
where $y^2=\sum_{i} \kappa^2$. We have confirmed numerically that Eq.\,\ref{eq:nu_SM_NT} is a reasonable check 
to keep the DS and SM neutrino out of thermal equilibrium for any temperature of the SM neutrino bath.

We start with the assumption that both $\phi$ and $S$ have zero initial abundance. The inverse 
decay of SM neutrino yields $\phi$ and conversion of $\phi$ takes place through $\phi\phi\rightarrow 
S S$ process. At a later stage when number density of $S$ is sufficient, the reverse 
process $SS\rightarrow \phi\phi$ turns effective (see Fig.\,\ref{fig:feyn_dia}). If the rate of such forward and reverse 
interactions in dark sector are fast enough, a dark equilibrium can be formed with a new 
temperature $T_D=T_S\neq T_\gamma,\,T_\nu$. Assuming the dark sector is internally thermalised,
{ we have estimated the dark sector temperature $T_D$ by solving the Boltzmann equation of total dark sector 
energy density $\rho_D=\rho_S+\rho_\phi$.} The equilibration of the dark sector can be ensured
by the following condition.
\begin{align}
 n_S^{\rm eq}(T_D)\langle \sigma v\rangle_{SS\rightarrow\phi\phi}> H, 
 \label{eq:DSeqO} 
\end{align}
where we have considered $\langle \sigma v\rangle_{SS\rightarrow\phi\phi}$ to be
$s$-wave dominated. Simultaneous requirements of dark sector equilibrium and non-thermalisation of
dark and visible sectors put constraints on the model parameters 
and the consequences of these constraints
will be discussed in section \ref{sec:BE}.

We are particularly interested in the dynamics of a MeV scale secluded dark sector in presence of 
dark $S-\phi$ equilibrium. The impact of such MeV scale dark sector on Planck can be 
parameterized by the effective number of relativistic species at CMB as defined by,
\begin{align}\label{eq:Neff}
 N_{\rm eff}^{\rm CMB} =\frac{8}{7}
\left(\frac{11}{4}\right)^{4/3}\left(\frac{\rho_\nu}{\rho_\gamma}\right),
\end{align}
where $\rho_\gamma=2\times\frac{\pi^2}{30}T_\gamma^4$, $\rho_\nu=2\times 3\times\frac{7}{8}
\times\frac{\pi^2}{30}T_\nu^4$ are the energy densities of photon and neutrino baths respectively.  
Before neutrino decoupling, the temperatures of the neutrino bath and the photon bath are equal. 
In the post neutrino decoupling era, $T_\gamma$ and $T_\nu$ evolve separately. If all the BSM 
particles are having mass much larger than eV scale, they do not contribute to the energy 
budget during CMB and hence in Eq.\,\ref{eq:Neff}, one can write,
\begin{align}
N_{\rm eff}^{\rm CMB}=3\times\left(\frac{11}{4}\right)^{4/3}\left(\frac{T_\nu}{T_\gamma}\right)^4.
\label{eq:NeffBBN}
\end{align}
Similarly one can define the number of relativistic degrees of freedom
at the time of BBN and it is given by ({ considering all BSM particles are non-relativistic during BBN})
\begin{align}
N_{\rm eff}^{\rm BBN} = \dfrac{8}{7}\left(\dfrac{\rho_\nu}{\rho_\gamma}\right).
\end{align}

\section{Boltzmann equations and Numerical analysis}
\label{sec:BE}
To study the effect of the dark sector dynamics on
$\Delta N_{\rm eff}$, we need to solve the Boltzmann 
equations for comoving number densities $Y_{i}$ ($i = \phi$, $S$), $\xi = T_\nu/T_\gamma$,
and $\xi_D = T_D/T_\gamma$. In deriving the 
above Boltzmann equations, we have used the following
assumptions.

\begin{itemize}
\item The temperatures of all the neutrino flavors
are same i.e. $T_{\nu_e} = T_{\nu_\mu} = T_{\nu_\tau} \equiv T_\nu$.
\item The entropy of the photon bath is conserved since at late time
$\phi$ can only decay into a pair of neutrinos. { Using the conservation of entropy in the photon bath, we have 
defined $Y_i=n_i/s_\gamma$ and the relic abundance of the DM is calculated accordingly\footnote{ We estimate the 
relic abundance of DM using $\Omega_S h^2=\frac{g_{*s,\gamma}}{g_{*s}^{\rm today}}\times 2.745
\times 10^8\times Y_S^{(x\rightarrow \infty)}\times \left(\frac{m_{S}}{1\,\text{GeV}}\right)$ where $g_{*s,\gamma}=2$ and $g_{*s}^{\rm today}=3.91$.}.}

\item The interaction rate between the dark sector particles such 
as $S$ and $\phi$ is sufficient enough to keep them in thermal equilibrium
and they share a common temperature $T_D$ 
(see discussion above Eq.\,\ref{eq:DSeqO}).
\end{itemize}

Using the above assumptions, first we write the evolution
of $\xi$ as a function of $x \equiv m_S/\tg$.
\begin{align}
\dfrac{d \xi} {d x} &= 
\dfrac{1}{x} \left[ 
\xi -\dfrac{\mathcal{B}_{T_\gamma}(T_\gamma, T_\nu)}{\mathcal{B}_{T_\nu} (T_\gamma, T_\nu)}
\right]\,\,\,,
\label{XI}
\end{align}
where 
\begin{align}
&\mathcal{B}_{\tn}(\tg, \tn)= \dfrac{\mathcal{C}_{\rm el} (\tg,\tn) + \mathcal{C}_{\rm inel} (\tg,\tn) - \mathcal{C}_{\nu \nu \lra \phi} - 
4 H \rho_\nu}{\frac{\partial \rho_\nu}{ \partial \tn}}\,\,,\nn\\
&\mathcal{B}_{\tg}(\tg, \tn)= \dfrac{-\mathcal{C}_{\rm el} (\tg,\tn) - \mathcal{C}_{\rm inel} (\tg,\tn) - 
4 H \rho_\gamma-3H(\rho_e + p_e)}{\frac{\partial (\rho_\gamma + \rho_e)}{ \partial \tg}}\,\,.\nn\\
\end{align}
Here, $\mathcal{C}_{\rm el} (\tg,\tn)$ and $\mathcal{C}_{\rm inel} (\tg,\tn)$ are the relevant collision terms for the
elastic and inelastic processes such as $e^+ e^- \lra \bar{\nu}_i \nu_i$,
$e^-(e^+) \nu_i (\bar{\nu}_i) \lra e^- (e^+) \nu_i (\bar{\nu}_i)$, $e^+(e^-) \nu_i (\bar{\nu}_i) \lra e^+ (e^-) \nu_i (\bar{\nu}_i)$.
$\mathcal{C}_{\nu \nu \lra \phi}$ is the collision term for $\phi \lra \bar{\nu}_i \nu_i$
where $i = e, \mu, \tau$. The matrix amplitude square of these processes and the generic form of 
collision terms for elastic, inelastic, and decay processes
are given in decay processes are given in table\,\ref{table1} of appendix\,\ref{sec:A2} and appendix\,\ref{sec:A3} respectively. The energy densities 
of electron, photon, and neutrino are respectively denoted by $\rho_e$, $\rho_\gamma$, and
$\rho_\nu$. $p_e$ denotes the pressure of electron.

The evolution of the co-moving number densities of $\phi$ ($Y_\phi$), and 
$S$ ($Y_S$) are governed by the following two coupled Boltzmann equations.

\begin{align}
\dfrac{d Y_\phi}{ d x} &= \dfrac{h_{\rm eff} s_\gamma}{x H}
\langle \sigma v_{\rm rel} \rangle_{SS \to \phi \phi}^{T_D}
\left[ Y_S^2 - \left(\dfrac{Y_S^{\rm eq}(T_D)}
{Y_\phi^{\rm eq}(T_D)}\right)^2 Y_\phi^2
\right] \nn \\
&+ \dfrac{h_{\rm eff}}{x H}\left(\langle \Gamma_{\phi} \rangle_{T_\nu}
Y_\phi^{\rm eq}(\tn) - \langle \Gamma_{\phi} \rangle_{T_D}
Y_\phi \right)\,\,\,
\label{Yphi}
\end{align}

\begin{align}
\dfrac{d Y_S}{ d x} &= -\dfrac{h_{\rm eff}s_\gamma}{x H}
\langle \sigma v_{\rm rel} \rangle_{SS \to \phi \phi}^{T_D}
\left[ 
Y_S^2 - \left(\dfrac{Y_S^{\rm eq}(T_D)}
{Y_\phi^{\rm eq}(T_D)}\right)^2  Y_\phi^2
\right]\,\,\,.
\label{YS}
\end{align}
Here, $s_\gamma$ is the entropy density of the photon bath, and $Y^{\rm eq}_{\phi (S)} (T_D)$ denotes the
equilibrium co-moving number density of $\phi (S)$ at hidden sector
temperature $T_D$. The Hubble parameter $H$ and $h_{\rm eff}$
are defined as follows.
\begin{align}
H &= \sqrt{\dfrac{8 \pi}{3 M^2_{\rm Pl}} \left(\rho_\gamma 
+ \rho_\nu+\rho_S+\rho_\phi\right)}\,\,,\nn\\
h_{\rm eff} &= 1 + \dfrac{1}{3}\dfrac{d \ln g_{*s,\gamma} 
(T_\gamma)}{d \ln T_\gamma}\,\,,
\end{align}
where $M_{\rm Pl} = 1.22 \times 10^{19}\,\rm GeV$ and
$g_{*s,\gamma}(T_\gamma)$ is the relativistic degrees of freedom
contributing to the entropy density of the photon bath. The energy densities of
photon and neutrino baths are indicated by
$\rho_\gamma$, $\rho_\nu$ respectively. { Note that, in the expression of the Hubble parameter, we have included 
the energy densities of dark sector particles in addition to the contributions from SM fields.}
On the right hand side (RHS)
of Eq.\,\ref{Yphi} and Eq.\,\ref{YS}, $\langle \sigma v_{\rm rel} 
\rangle^{T_D}_{SS\to \phi \phi}$
stands for the thermally averaged annihilation cross section of $SS\to \phi \phi$ at temperature
$T_D$. The analytical form of $\sigma_{SS\to \phi \phi}$ is given by,
\bea\label{eq:cross-sec}
\sigma_{SS\to \phi \phi} = \dfrac{\lambda^2}{32 \pi s} \sqrt{\dfrac{s-4m_\phi^2}{s-4m_S^2}}\,\,,
\eea
where $\sqrt{s}$ is the centre of mass energy of the collision. $\langle \Gamma_{\phi} \rangle _{T_D}$ and 
$\langle \Gamma_{\phi} \rangle _{T_\nu}$ on the RHS of Eq.\,\ref{Yphi}
are the thermally averaged total decay width of $\phi$ at 
temperature $T_D$ and $T_\nu$ respectively.
A general expression of the thermally averaged decay width at temperature
$T_i$ is given by $\langle \Gamma_\phi \rangle _{T_i} = \Gamma_\phi 
\dfrac{K_1(m_\phi/T_i)}{K_2 (m_\phi/T_i)}$, where
$\Gamma_\phi=\frac{y^2 m_\phi}{16\pi}$ is the total decay width of $\phi$. In deriving the Boltzmann 
equations for $Y_\phi$ and $Y_\chi$, we have also taken into account the effect of non-zero chemical potential for 
dark sector particles as evident from the presence of square of the ratio $(Y_S^{\rm eq}(T_D)/Y_\phi^{\rm eq}(T_D))$ in the RHS of Eq.\,\ref{Yphi}.

As we discussed earlier, the dark sector is connected to SM by
the portal coupling $y$ and our choice of $y$ keeps dark and the visible sector out of thermal equilibrium.
{Initially inverse decay of $\phi$ populates the dark sector and subsequently $\phi\phi\to SS $ process occurs. 
The Boltzmann equation that governs the evolution of total dark sector energy density ($\rho_D=\rho_\phi+\rho_S$) is read as
\begin{align}
\frac{d\rho_D}{dt}+3 H(\rho_D+p_D)=\mathcal{C}_{\nu\nu\leftrightarrow \phi},
\end{align}
where $p_D$ stands for the pressure and $\mathcal{C}_{\nu\nu\leftrightarrow \phi}$ includes the collision integrals for 
both decay and inverse decay processes of $\phi$. 
We use the assumption of Maxwell- Boltzmann statistics and write $\rho_D$ and $p_D$ as function of $T_D$ as given by 

\begin{align}
\vspace{1mm}
&\rho_D=\left(\frac{Y_\phi}{Y_\phi^{\rm eq}(T_D)}\right)\rho_\phi^{\rm eq}(T_D)+\left(\frac{Y_S}{Y_S^{\rm eq}(T_D)}\right)\rho_S^{\rm eq}(T_D),\label{eq:rhoD}
\end{align}
\begin{align}
&p_D=\left(\frac{Y_\phi}{Y_\phi^{\rm eq}(T_D)}\right)p_\phi^{\rm eq}(T_D)+\left(\frac{Y_S}{Y_S^{\rm eq}(T_D)}\right)p_S^{\rm eq}(T_D),\label{eq:pD}
\end{align}
In the RHS of the above equation, the quantities, $\rho_\phi^{\rm eq}$ and $p_\phi^{\rm eq}$ represent the energy density and pressure at equilibrium as given by
\begin{align}
&\rho_i^{\rm eq}(T_D)=\frac{g_i}{2\pi^2}T_D^4\left[3\left(\frac{m_i}{T_D}\right)^2K_2\left(\frac{m_i}{T_D}\right)+
\left(\frac{m_i}{T_D}\right)^3K_1\left(\frac{m_i}{T_D}\right)\right],\nonumber\\
&p_i^{\rm eq}=\frac{g_i}{2\pi^2}T_D^4\left(\frac{m_i}{T_D}\right)^2K_2\left(\frac{m_i}{T_D}\right),
\end{align}
where once again, the ratio $\left(\frac{Y_i}{Y_i^{\rm eq}(T_D)}\right)$ in Eqs.\ref{eq:rhoD}-\ref{eq:pD} takes care of 
the non-zero chemical potential during the temperature evolution for the i$^{\rm th}$ species. Utilizing these, we obtain,
\begin{align}
 \frac{dT_D}{dx}=\frac{s_\gamma Y_\phi^{\rm eq}(T_D)Y_S^{\rm eq}(T_D)\left\{\frac{1}{s_\gamma}\mathcal{C}_{\nu\nu\leftrightarrow\phi}-
 3 H(Y_\phi+Y_S)T_D\right\}+F_1+F_2}{x H\sum_{i\neq j}Y_iY_j^{\rm eq}\left\{\frac{d\rho_{i}^{\rm eq}}{d T_D}-\frac{1}{n_i^{\rm eq}(T_D)}
 \frac{dn_i^{\rm eq}(T_D)}{d T_D}\rho_i^{\rm eq}(T_D)\right\}},\label{eq:DT}
\end{align}
where, 
\begin{align}
& F_1=s_\gamma\langle\sigma v\rangle_{SS\to \phi\phi}^{T_D}\left[Y_S^2-\left(\frac{Y_S^{\rm eq}(T_D)}
{Y_\phi^{\rm eq}(T_D)}\right)^2Y_\phi^2\right]\left(\rho_S^{\rm eq}Y_\phi^{\rm eq}-\rho_\phi^{\rm eq}
Y_S^{\rm eq}\right),\nonumber\\
& F_2=-\rho_\phi^{\rm eq}(T_D)Y_\chi^{\rm eq}(T_D)\left[\langle\Gamma_{\phi\to \nu\nu}\rangle_{T_\nu}Y_\phi^{\rm eq}(T_\nu)-
\langle\Gamma_{\phi\to \nu\nu}\rangle_{T_D}Y_\phi\right]\nonumber.
\end{align}
\noindent and the indices $i$ and $j$ stand for $\phi$ and $S$.}
\begin{figure*}
\centering
\subfigure[\label{XI_phi_BP}]{\includegraphics[height = 6.5cm, width = 7.55cm]{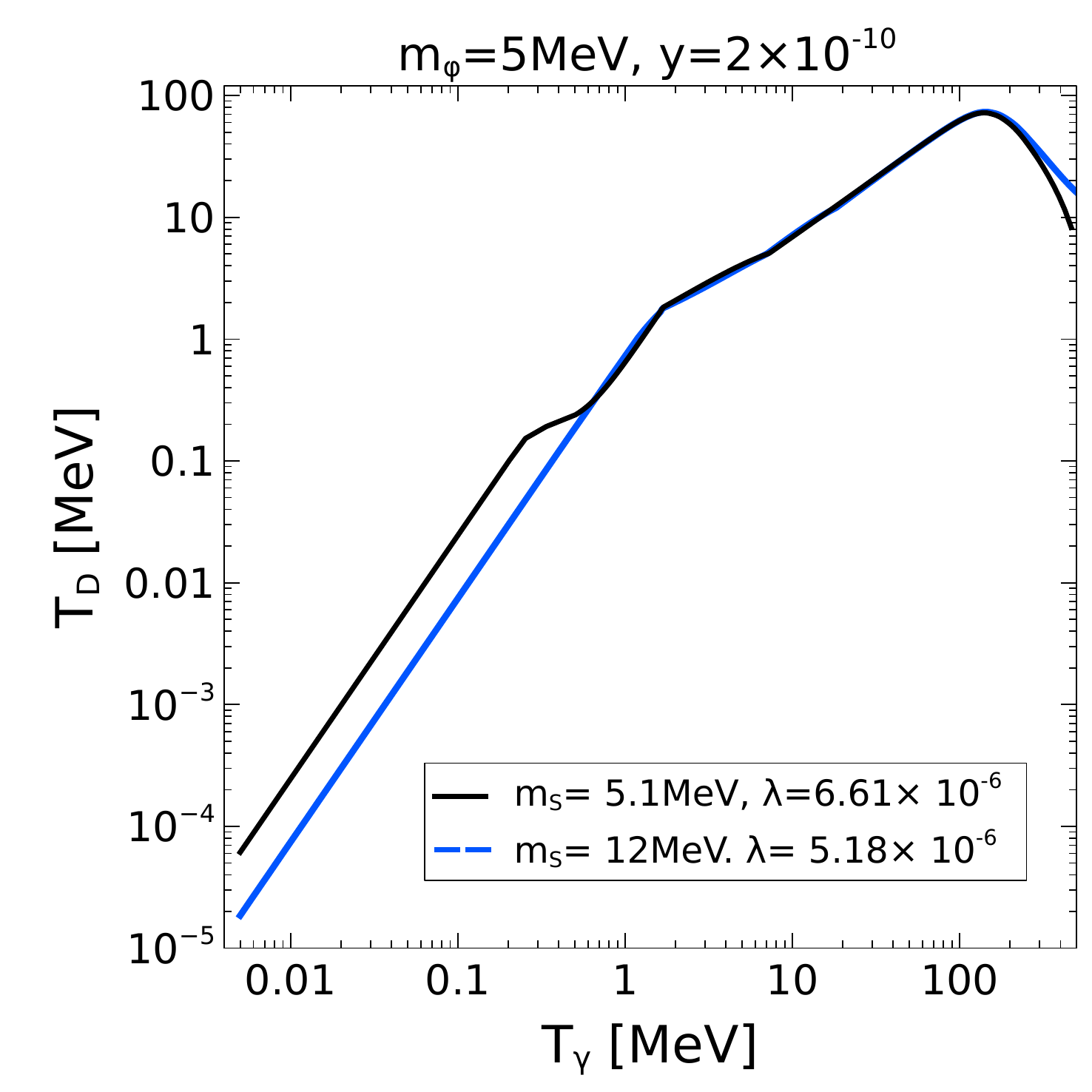}}
\subfigure[\label{relic_BP}]{\includegraphics[height=6.5cm,width = 7.55cm]{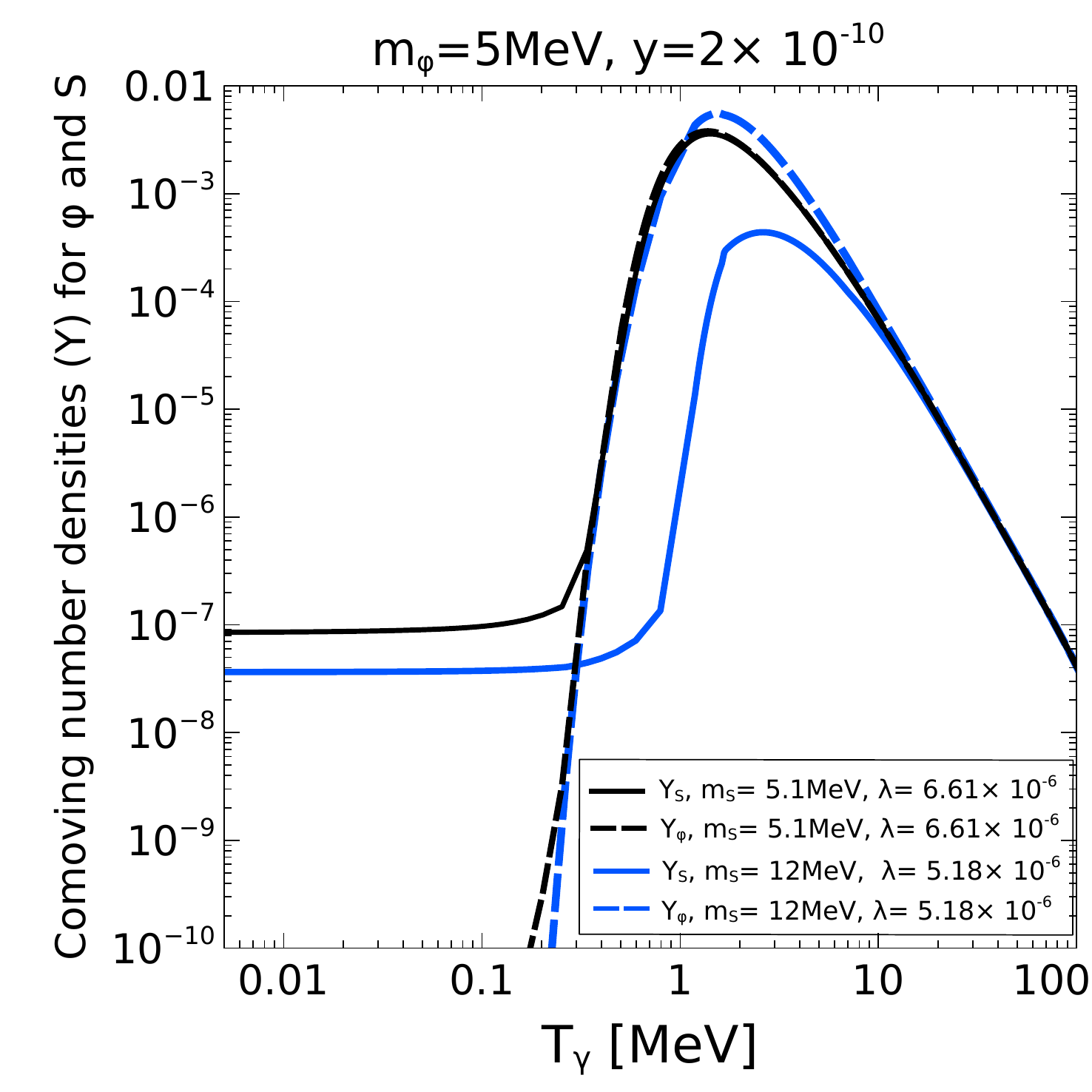}}
\subfigure[\label{delta_Neff_BP}]{\includegraphics[height=6.3cm,width = 7.8cm]{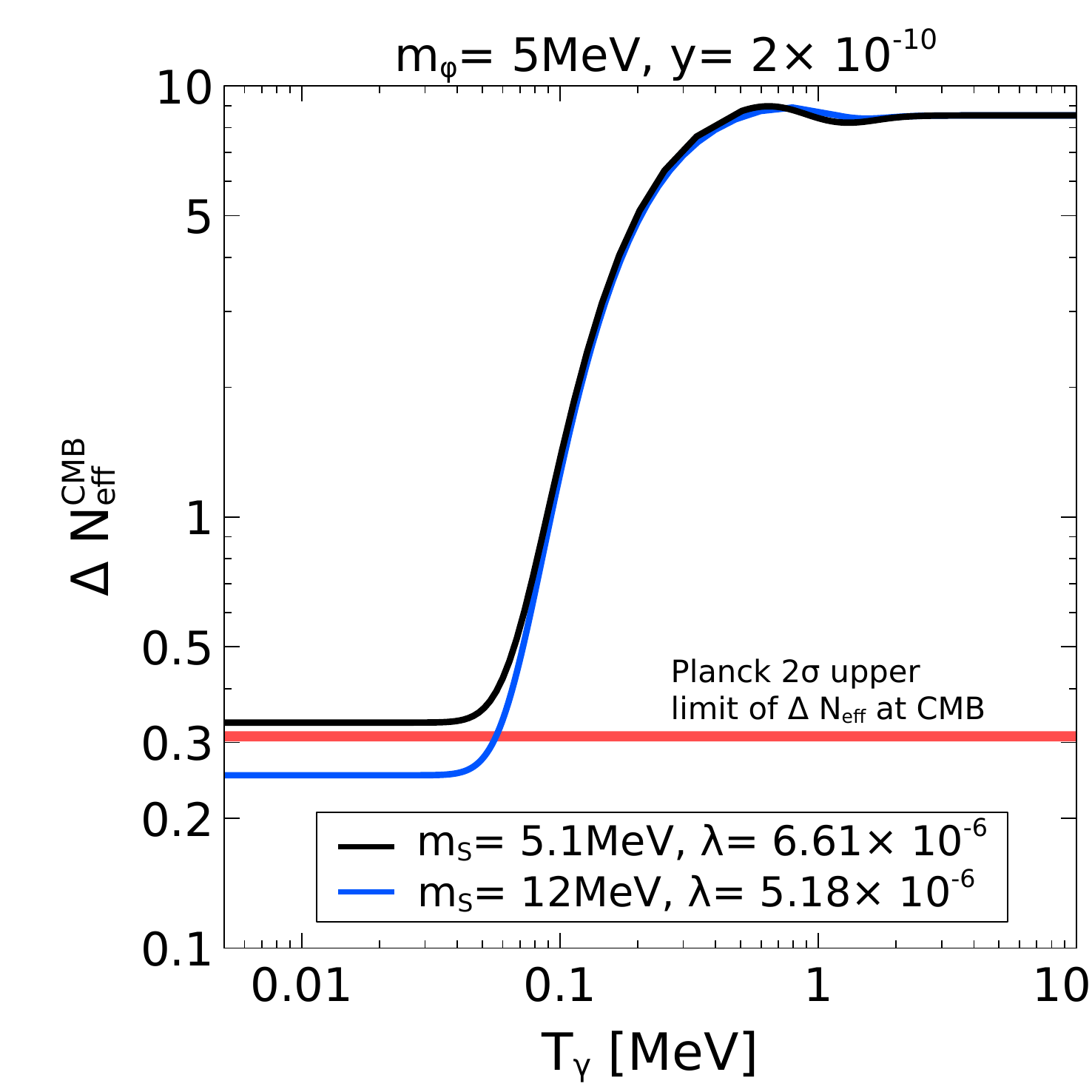}} 
\caption{(a) Variation of ${T_D}$ as a function of $\tg$
for $m_S = 5.1\rm MeV$, $\lambda = 6.61\times 10^{-6}$ (black line) and $m_S = 12\rm MeV$, 
$\lambda = 5.18 \times 10^{-6}$ (blue line) and these parameters yield the correct DM relic abundance \cite{Planck:2018vyg}.
(b) Evolution of co-moving number density of $\phi$ ($Y_\phi$) (dashed lines) 
and $S$ ($Y_S$) (solid lines) as a function of the SM bath temperature $\tg$ for the same choice of parameters as in
\ref{XI_phi_BP}.
(c)
The variation of $\Delta N^{\rm CMB}_{\rm eff}$ as a function of $\tg$. Here the parameter choices and color codes 
are same as Fig.\,\ref{XI_phi_BP}. The red horizontal solid line indicates the upper limit on $\Delta N_{\rm eff}$ at CMB by Planck $2\sigma$ data. In all plots, 
we have considered $m_\phi = 5\rm MeV$ and $y =2\times 10^{-10}$.}
\label{BP_plot}
\end{figure*}

\begin{figure*}
\centering
\subfigure[\label{fig:DN_VS_lam_mphi_2p85MeV}]{\includegraphics[height = 6.5cm, width = 7.2cm]
{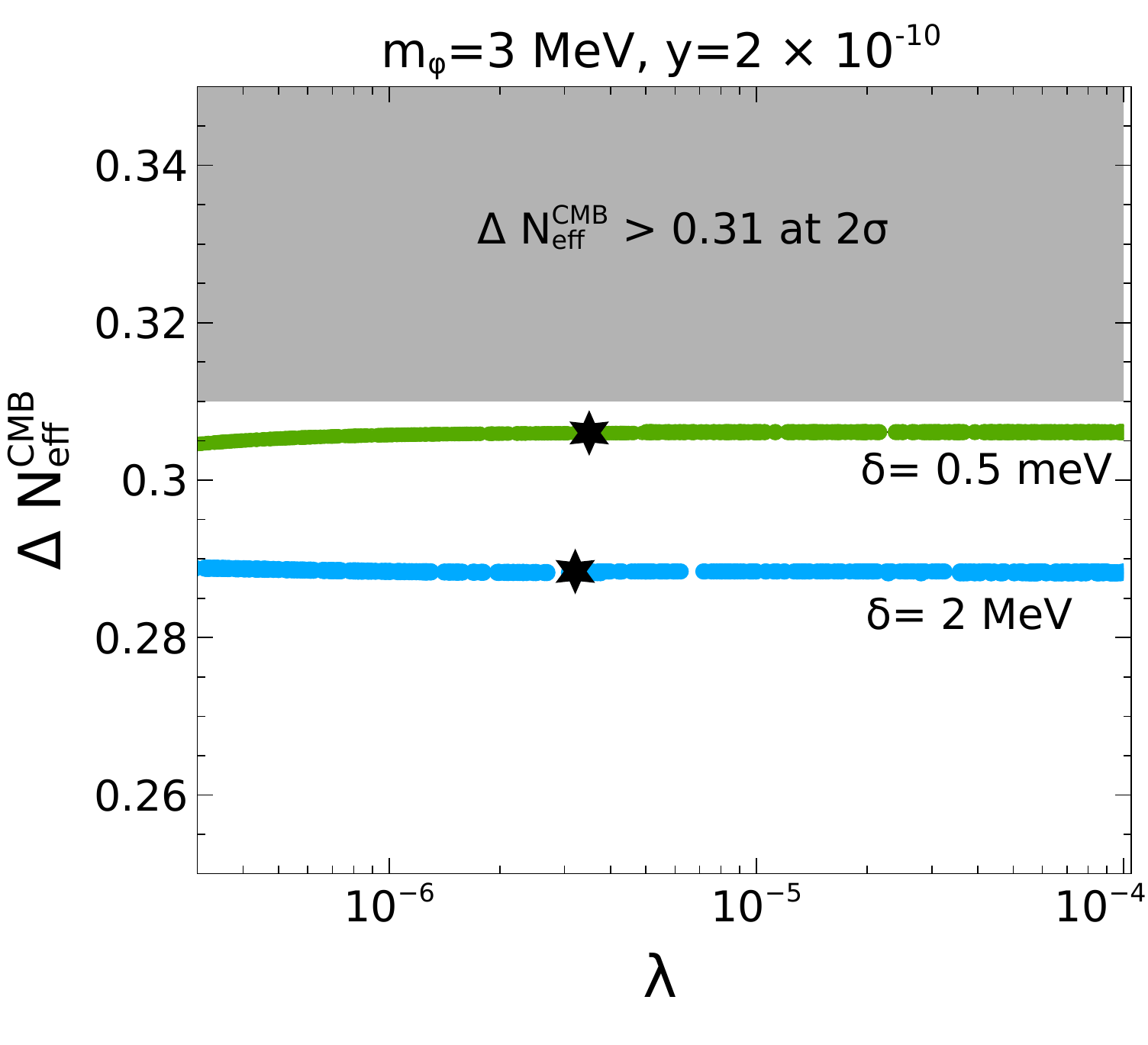}}~~~~
\subfigure[\label{fig:DN_VS_lam_mphi_5MeV}]{\includegraphics[height = 6.5cm, width = 7.2cm]
{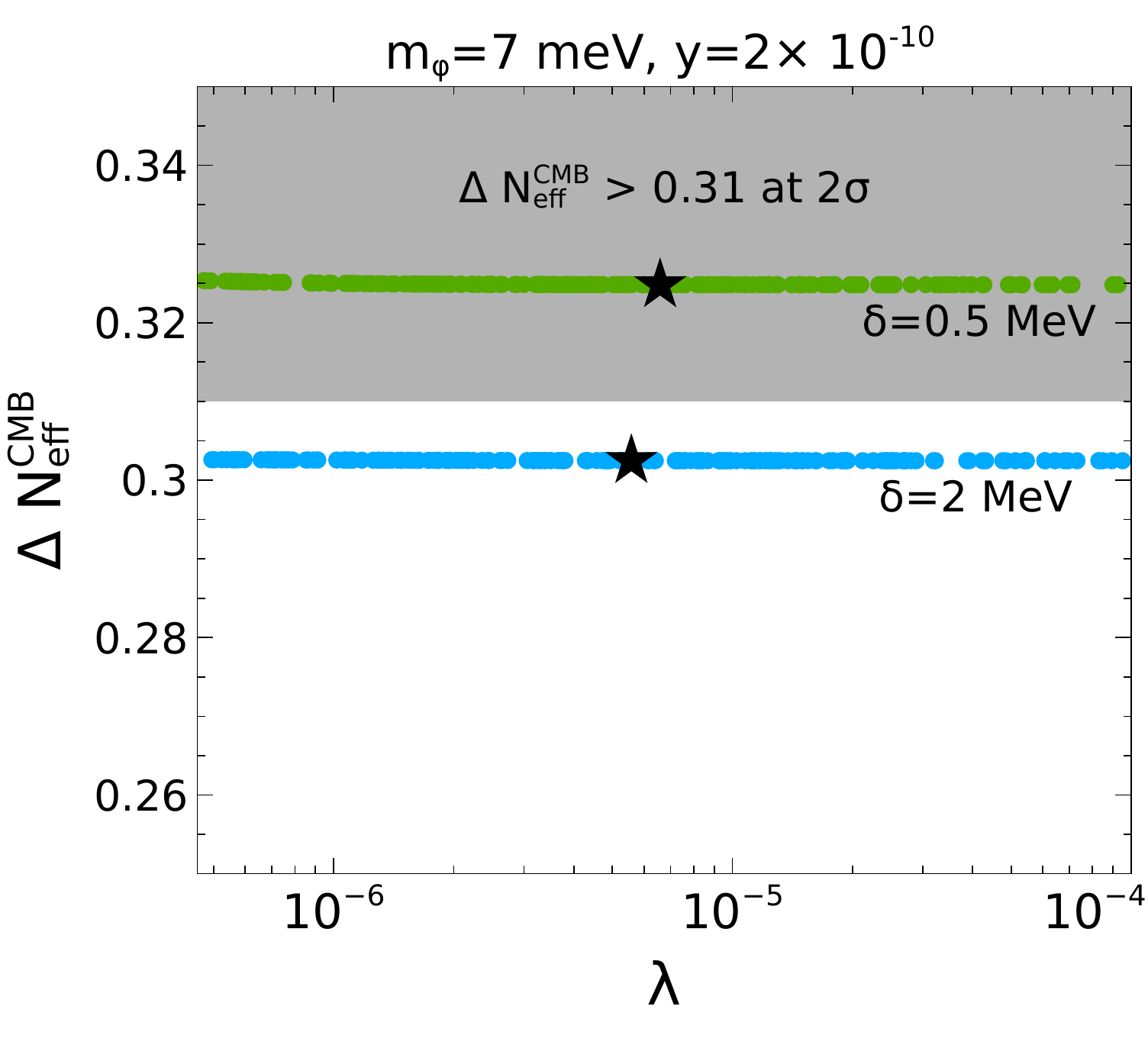}}
\subfigure[\label{fig:DN_VS_lam_mphi_10MeV}]{\includegraphics[height = 6.5cm, width = 7.2cm]
{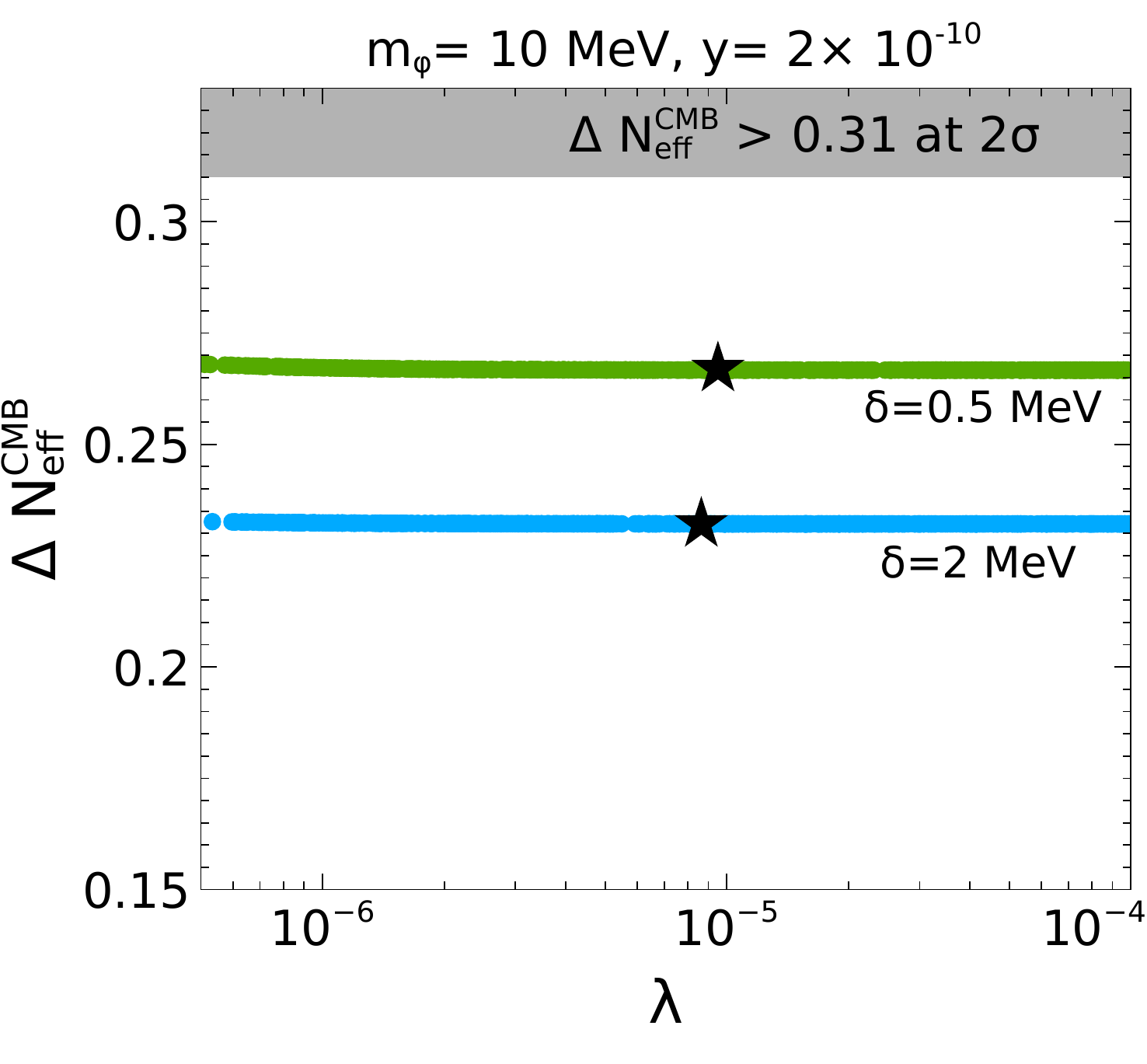}}  
\caption{$\Delta N^{\rm CMB}_{\rm eff}$ as a function of $\lambda$ for (a) $m_\phi=3$MeV, 
(b) $m_\phi=7$MeV and (c) $m_\phi=10$MeV. The green points are for $\delta=(m_S-m_\phi)=0.5$MeV 
whereas blue points indicate the $\Delta N_{\rm eff}^{\rm CMB}$ values 
for $\delta=2$MeV. Here we have considered $y = 2\times 10^{-10}$. 
The symbol `$\star$' stands for the $\lambda$ value that yields correct relic abundance \cite{Planck:2018vyg}. 
In all three figures, the grey region is disallowed from the 
measurement of $\Delta N^{\rm CMB}_{\rm eff}$ by Planck
within $2\sigma$ limit.}
\label{fig:DN_VS_lam_mchi}
\end{figure*}

At first, we would like to calculate the value of $\Delta N_{\rm eff}^{\rm CMB}$ in SM. In order to do so, we have 
numerically solved Eq.\,\ref{XI} in the absence of new physics. Following \cite{Hannestad:1995rs,Dolgov:2002wy,Kreisch:2019yzn},
we have computed the collision terms for SM $\nu$ decoupling, considering Fermi-Dirac (FD)
distribution function for all the SM fermions and appropriate Pauli blocking factors. For simplification, we have neglected 
the mass of the electron ($m_e$) since at the time of $\nu$
decoupling, $\tg \gg m_e$. We have obtained the value of $\tn/\tg = 0.714$ at CMB and the correspondingly 
$N_{\rm eff}^{\rm CMB, SM}\simeq 3.02$. Now we proceed to estimate $N_{\rm eff}^{\rm CMB}$ in presence of the MeV 
scale secluded dark sector. We solve the coupled system of Boltzmann equations (Eqs.\ref{XI}-\ref{YS},\ref{eq:DT}) 
with the assumption of initial zero abundance for the dark sector particles and $T_D^{\rm initial}=0$ to study the dark 
sector dynamics along with the estimate of $\Delta {\rm N}_{\rm eff}^{\rm CMB}$. The obtained numerical 
results in presence of the new physics are presented as follows.

\begin{figure}
\begin{center}
\includegraphics[height = 8.5cm, width = 10cm]{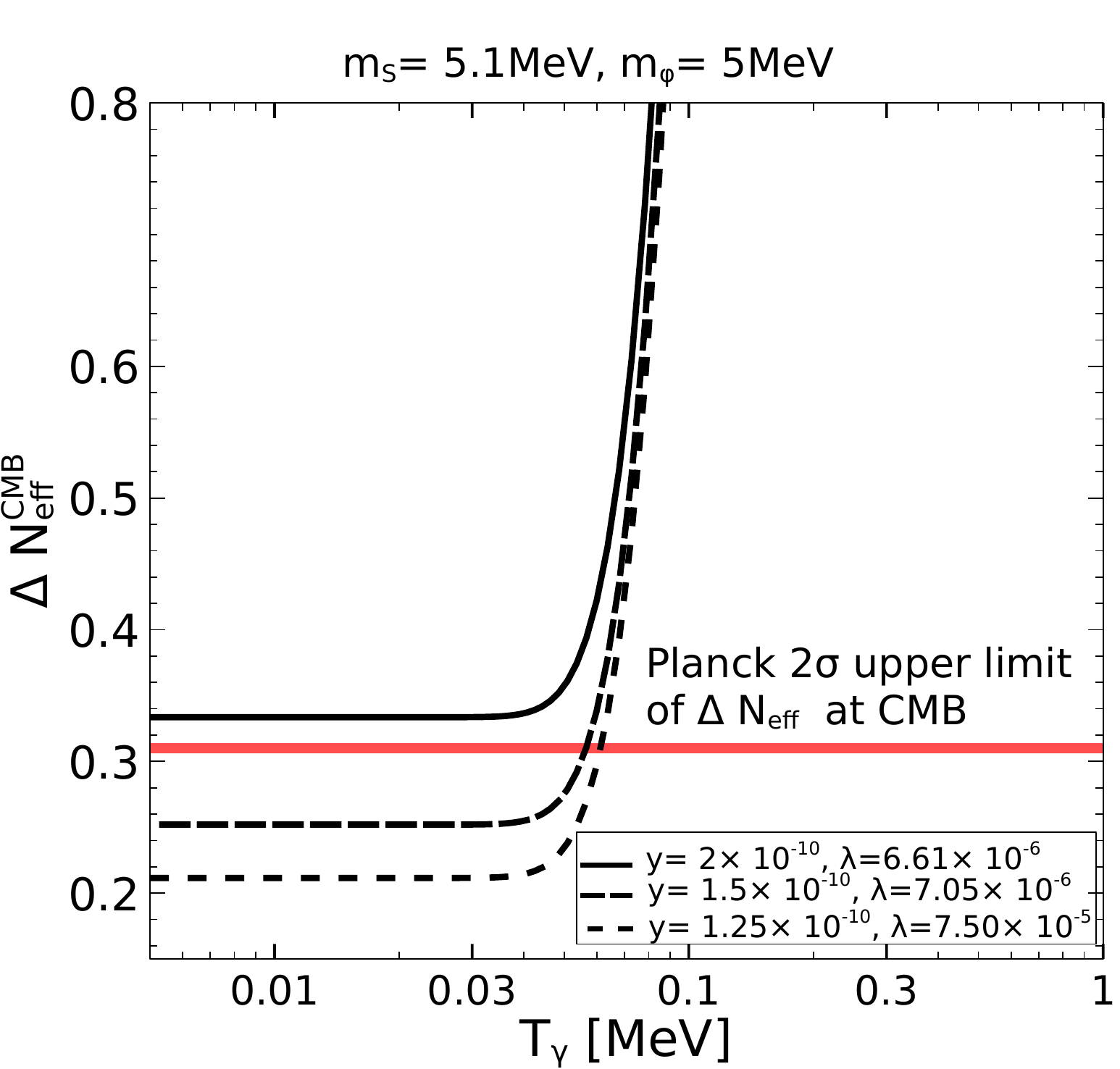}
\end{center}
\caption{The variation of $\Delta N^{\rm CMB}_{\rm eff}$
as a function of $\tg$ for $y = 2\times 10^{-10}$ (solid line), $y = 1.5 \times 10^{-10}$ (dashed line),
and $y = 1.25 \times 10^{-10}$ (dotted line). Here we choose $m_S = 5.1\rm MeV$, $m_\phi = 5\rm MeV$. The red 
horizontal solid line indicates the upper limit on $\Delta N_{\rm eff}$ at CMB by Planck $2\sigma$ data.}
\label{fig:DN_VS_T_y}
\end{figure}
In Fig.\,\ref{BP_plot}, we demonstrate the evolution patterns of $T_D$, $Y_\phi$, $Y_S$, and
$\Delta N_{\rm eff}^{\rm CMB}$ as functions of the SM temperature $\tg$. { We have fixed $m_\phi$ and $y$ at $5$MeV and 
$2\times 10^{-10}$ respectively and considered two specific DM masses which are 5.1MeV and 12MeV. The corresponding $\lambda$ 
value for each of the DM mass is fixed by the relic density requirement ($\Omega_S h^2\sim 0.12$).
The dark sector temperature starts from zero initial value and increases at the early stage of its evolution due to the 
production of $\phi$ from $\bar{\nu}_i \nu_i \to \phi$ process. The temperature $T_D$ continues to grow until it reaches a 
maximum value, after which it starts to redshift like radiation upto $T_D \lesssim m_\phi$ due to the expansion of the 
Universe. Around $T_\gamma\sim m_\phi$, we notice that the redshift for $T_D$ gets slower. This occurs since the 
production rate of $\nu\nu\rightarrow \phi$ reaches maximum around this temperature. For $m_S=5.1$MeV we also observe 
the impact of dark matter freeze out in the evolution of $T_D$, where late time conversion process $SS\to \phi\phi$ 
slows down the redshift of $T_D$ for the second time. This pattern is not visible for $m_S=12$MeV as in this case the 
dark matter freeze out occurs at a higher temperature when the $\nu\nu\to\phi$ process is still active. Finally after 
the freeze out of the dark matter and decay of $\phi$, the dark sector temperature redishifts like a non-relativistic 
matter field with usual scale factor ($a$) dependence $T_D\propto a^{-2}$.}  

In Fig.\,\ref{relic_BP}, we show the evolutions of $Y_S$ (solid lines) and $Y_\phi$ (dashed lines) as functions
of $\tg$ considering two different choices of $(m_S,\lambda)$ as indicated by black and blue curves for $m_S=5.1$MeV and 12MeV, 
respectively. The corresponding values of $\lambda$ are chosen accordingly to yield correct relic abundance with a larger 
DM mass requiring lower value of $\lambda$ to obey the relic density bound. On the other hand, in Fig.\,\ref{delta_Neff_BP}, 
we show the variation of $\Delta N_{\rm eff}^{\rm CMB}$ as a function of $\tg$ for the same choice of parameters
as considered in Fig.\,\ref{relic_BP}.
It is observed from Fig.\,\ref{BP_plot} that the dark matter $S$ of mass $12\rm MeV$ freezes-out much earlier and at the 
time of DM freeze-out, the production of $\phi$ is still in progress. As a result, when $\phi$ starts decaying to a pair of 
SM neutrinos, the DM has already decoupled
from $\phi$ bath and the final value of $\Delta N_{\rm eff}$ is governed mostly by
$m_\phi$ and $y$. However the situation turns a bit different as we decrease the mass hierarchy
of $S$ and $\phi$. In case of $m_S = 5.1\rm MeV$, decoupling of $S$ is delayed in comparison to the earlier scenario and
at the time of decay of $\phi$, $SS\to \phi \phi$ is still active with $S$ having equivalent comoving abundance of 
$\phi$ around $T_D\sim m_\phi$. As a result, the depletion
of $\phi$ has been counterbalanced by the production $SS\to \phi \phi$, which in turn
enhances $\Delta N_{\rm eff}^{\rm CMB}$. 
Thus, $\Delta N_{\rm eff}^{\rm CMB}$ increases as we reduce the mass gap
between $S$ and $\phi$.
 
Next, the dependence of the parameter $\lambda$ on $\Delta N_{\rm eff}^{\rm CMB}$ is shown
in Fig.\,\ref{fig:DN_VS_lam_mchi} for $y=2\times 10^{-10}$ and three different values of $m_\phi$. The lowest 
chosen value of $m_\phi$ is 3MeV in Fig.\,\ref{fig:DN_VS_lam_mchi} and for this choice, $\phi$ remains out of 
equilibrium from SM neutrino bath as can be confirmed from Eq.\,\ref{eq:nu_SM_NT}. For larger $m_\phi$ 
with same $y$, the Eq.\,\ref{eq:nu_SM_NT} is easier to satisfy. 
We consider two different hierarchical patterns of dark sector particles 
which are $(m_S-m_\phi)$= 0.5MeV and 2MeV. Let us define $\delta=(m_S-m_\phi)$ for 
convenience. The dark sector reaches internal equilibrium for the chosen ranges of $\lambda$ in Figs.\,\ref{fig:DN_VS_lam_mphi_2p85MeV}-\ref{fig:DN_VS_lam_mphi_10MeV}
and it has been numerically checked. We find that $\Delta N_{\rm eff}^{\rm CMB}$ is insensitive to $\lambda$. 
This is because when the dark sector is internally thermalized, 
the abundance of $\phi$ at the onset of its decay is mostly governed by $y,m_\phi$ and $m_S$.
We also notice $\Delta N_{\rm eff}$ increases 
with the decrease in mass gap between $S$ and $\phi$ as earlier observed in Fig.\,\ref{delta_Neff_BP}. In all
three subfigures of Fig.\,\ref{fig:DN_VS_lam_mchi}, the point highlighted by `$\star$' 
indicates the required $\lambda$ to yield correct relic density for DM \cite{Planck:2018vyg}. Another
important outcome of Fig.\,\ref{fig:DN_VS_lam_mchi} is related to the dependence of
$m_\phi$ on $\Delta N_{\rm eff}^{\rm CMB}$. When $m_\phi$ is fixed at 3MeV, the predictions for $\Delta
N_{\rm eff}^{\rm CMB}$ considering both the
mass differences $\delta=$ 0.5MeV and 2MeV are below the Planck limit. With the increase of $m_\phi\sim 7$MeV with $\delta=0.5$MeV, the 
prediction for $\Delta N_{\rm eff}^{\rm CMB}$ enters into the disallowed region. This happens since larger $m_\phi$ increase 
the decay width of $\phi$ and causes enhanced rate of entropy injection into the SM neutrino bath. Further increase of 
$m_\phi$ to 10MeV shows different pattern and $\delta=0.5$MeV again makes a comeback to the allowed region of 
$\Delta N_{\rm eff}^{\rm CMB}$. This is due to the fact that a sufficient large $m_\phi$ would make $\phi$ to decay early 
with less impact on evolution of $T_\nu$. On the other hand for $\delta=2$MeV, the predictions for $\Delta N_{\rm eff}^{\rm CMB}$ 
always remain inside the Planck favored region.

 As we find from Fig.\,\ref{delta_Neff_BP}, the predictions on $\Delta
N_{\rm eff}^{\rm CMB}$ for $m_\phi=5$MeV and $m_S=5.1$MeV are disfavored when $y=2\times 10^{-10}$. Now, let us examine the 
effects of varying $y$ for the same set of $m_\phi$ and $m_S$. 
 In Fig.\,\ref{fig:DN_VS_T_y}, we show the variation of $\Delta N^{\rm CMB}_{\rm eff}$
as a function of $\tg$ for three different choices of the portal coupling $y$ by fixing
$m_\phi$ and $m_S$ at 5MeV and 5.1MeV respectively. 
The choices for $\lambda$ are made to obtain the best fit value of observed relic abundance $\Omega
h^2= 0.1200\pm 0.0012$ \cite{Planck:2018vyg} after formation of internal dark equilibrium.
Earlier we found that $y = 2 \times 10^{-10}$ is ruled out by present CMB data which is observed once 
again in Fig.\,\ref{fig:DN_VS_T_y}. However reducing $y$ suppresses  
the energy injection rate and thus decreases $\Delta N^{\rm CMB}_{\rm eff}$ and hence would be allowed by the Planck data.

In Fig.\,\ref{fig:BBNPrediction}, we have shown the evolutions of $N_{\rm eff}^{\rm BBN}$ following the definition of 
Eq.\,\ref{eq:NeffBBN} considering three different values of $m_\phi$ with the $m_S$ is fixed by $m_S=m_\phi+0.5$MeV and $y=2\times 10^{-10}$. 
The magnitude of $\lambda$ is determined by the requirement of obeying the relic density bound. 
It turns out that $N_{\rm eff}^{\rm BBN}$ at $T_{\rm BBN}\sim 1$MeV increases with $m_\phi$ however stays inside the allowed 
$2\sigma$ limit \cite{Fields:2019pfx}. Importantly, for lighter $m_\phi$, $N_{\rm eff}^{\rm BBN}$ is smaller than its SM value 
which is approximately three. This owes to the fact that for the lightest $m_\phi =3$MeV, as considered here, the production of 
$\phi$ continues even during BBN from the inverse decay of $\phi$. This reduces the temperature of the SM neutrino bath at BBN and 
we observe a dip in the value of $N_{\rm eff}^{\rm BBN}$ at $T_{\rm BBN}\sim 1$MeV for $m_\phi=3$MeV. {Overall, we find that 
BBN requirements do not impose any serious constraints to our model parameter space. A detailed analysis in order to study the 
impact of a secluded non-minimal dark sector on the  abundances of the primordial light elements ({\it e.g.} Helium-4 and 
Deuterium \cite{Arbey:2018zfh,An:2022sva}) is beyond the scope of the present article and will be pursued elsewhere.}

\begin{figure}[htb]
\vspace{0.5cm}
\begin{center}
\includegraphics[height =9cm, width = 11.5cm]{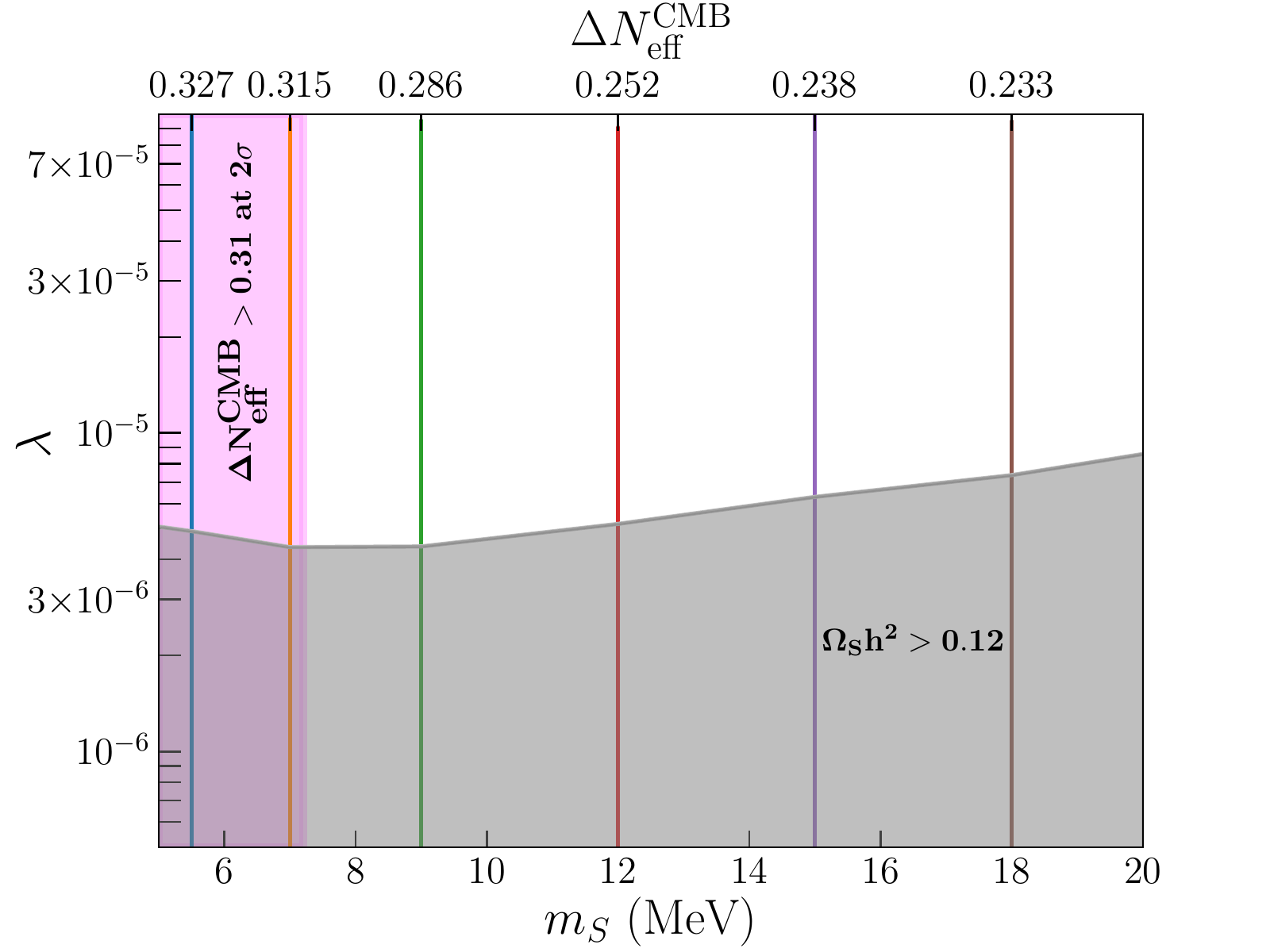}
\end{center}
\caption{ Allowed parameter space in $m_S - \lambda$ plane from the relic density bound along
with the $\Delta N^{\rm CMB}_{\rm eff}$ predictions for $m_\phi = 5 \rm MeV$ and $y = 2\times 10^{-10}$. The
overabundant region of DM is shown by grey shade. In the top x-axis, we show the predictions for $\Delta N_{\rm eff}^{\rm CMB}$ as function of DM mass $m_S$.}
\label{fig_ms_lambda_1}
\end{figure}

{ Finally in Fig.\,\ref{fig_ms_lambda_1}, we show the predictions for $\Delta N_{\rm eff}^{\rm CMB}$ in $m_S-\lambda$ plane. 
We have fixed $m_\phi$ and $y$ at 5MeV and $2\times 10^{-10}$ respectively. These particular choices ensure that $\phi$ does 
not equilibrate with SM neutrinos as followed from Eq.\,\ref{eq:nu_SM_NT}.
The overabundant region of DM is highlighted with light grey color.
We have numerically found that the condition for dark sector equilibrium following Eq.\,\ref{eq:DSeqO} with $m_{S}\lesssim 20$MeV 
poses much weaker constraint compared to the one from observed DM relic in the $m_S-\lambda$ plane. Hence we do not show it in Fig.\,\ref{fig_ms_lambda_1}. 
\begin{figure}[h]  
\begin{center}
\includegraphics[height = 8.5cm, width = 10cm]{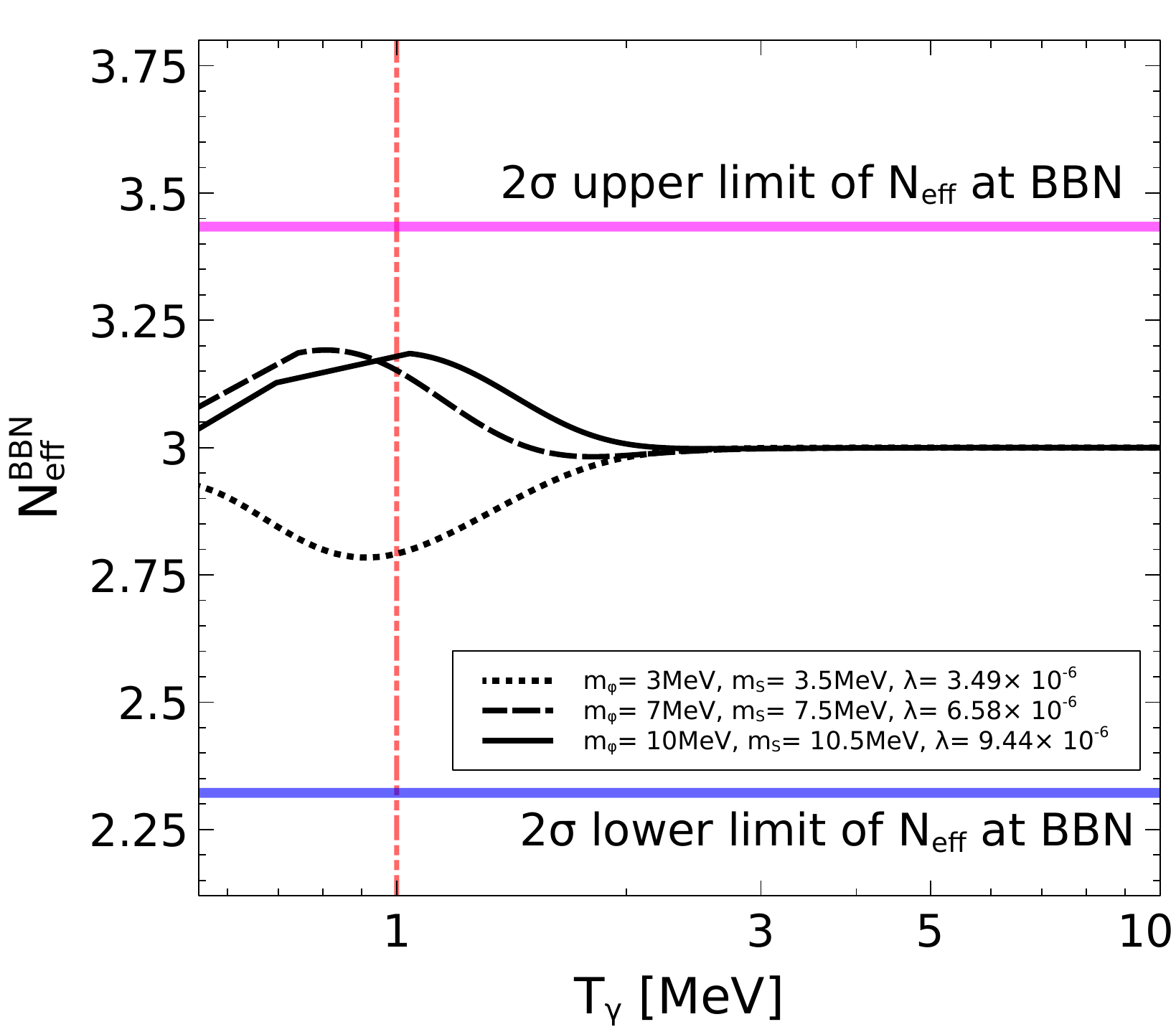}
\end{center}
\caption{ The evolution of $ N_{\rm eff}^{\rm BBN}$
with $T_\gamma$ is shown for three different sets of $(m_\phi,m_s)$ with $y=2\times 10^{-10}$. 
The corresponding $\lambda$ value for each set is fixed by the relic density requirement. 
The magenta and blue horizontal solid lines denote the upper and lower limit of $\Delta N_{\rm eff}$ at the time of BBN  at 
$2\sigma$ \cite{Fields:2019pfx} respectively. 
The perpendicular dot-dashed line represents $T_{\rm BBN}=1$MeV. }
\label{fig:BBNPrediction}
\end{figure}
The $\Delta N_{\rm eff}^{\rm CMB}$ lines for different $m_S$ are also depicted in Fig.\,\ref{fig_ms_lambda_1}. As explained 
earlier the predictions for $\Delta N_{\rm eff}^{\rm CMB}$ remain insensitive to variation in $\lambda$.
Clearly, one can see that the value
of $\Delta N_{\rm eff}^{\rm CMB}$ is maximum when $m_S$ turns closer to $m_\phi$. We also observed that as $m_S$ turns larger, 
its impact on $\Delta N_{\rm eff}^{\rm CMB}$ gets reduced. The latest Planck 2$\sigma$ bound on $\Delta N_{\rm eff}^{\rm CMB}$ 
restricts $m_S$ to be larger than around 7 MeV for $m_\phi=5$MeV and $y=2\times 10^{-10}$.}

\section{Summary and Conclusion}
In this work, we examine the scope of probing a MeV scale secluded dark sector from the 
observation of $\Delta N_{\rm eff}^{\rm CMB}$ by Planck. For the purpose, we have considered 
a simple structure of secluded dark sector comprising of two SM gauge singlet scalars $S$ 
and $\phi$. The dark sector never reaches thermal equilibrium with the SM sector.
The lighter scalar $\phi$ is neutrinophilic and we identify the heavier one $S$ as 
our DM candidate. Initially the dark sector particle $\phi$ gets populated from the SM sector 
by the $\overline{\nu}_i \nu_i \to \phi$ process and afterwards $\phi\phi\rightarrow SS$ process produces $S$. 
Depending on the interaction strength between $S$ and $\phi$, dark thermalisation can occur which 
causes freeze-out of $S$ via $SS\to \phi \phi$ process. We constrain our parameter space from the 
observations of both dark matter relic density and measurement of $N_{\rm eff}^{\rm CMB}$ by Planck ensuring internal dark thermal equilibrium. 
We particularly emphasize the impact of the dark sector parameters on the prediction 
of $N_{\rm eff}^{\rm CMB}$. Below we summarize few important observations that came out from our analysis.

$\bullet$ { We obtain unique predictions for the $\Delta N_{\rm eff}^{\rm CMB}$ for a particular DM mass that satisfies the 
relic density bound with suitable value of $\lambda$, provided mass of the other scalar and its coupling strength with the SM neutrinos are fixed.}

$\bullet$ The dark matter mass has a nontrivial role on the prediction of $\Delta N_{\rm eff}^{\rm CMB}$. The predictions for 
$\Delta {N_{\rm eff}^{\rm CMB}}$ turn out to be maximum in the nearly degenerate 
spectrum of dark sector particles and consideration of an increased hierarchy between the 
dark sector particles reduces the impact of dark matter mass on $\Delta N_{\rm eff}^{\rm CMB}$. We also find that prediction for 
$\Delta N_{\rm eff}^{\rm CMB}$ is insensitive to the dark sector interaction rate, provided the dark thermal equilibrium is reached.

$\bullet$ Increasing $m_\phi$ upto a certain value keeping other parameters ($y,~\lambda$ and $\delta$) fixed, 
enhances $\Delta {N_{\rm eff}^{\rm CMB}}$ hence restricted by the present Planck limit. When $m_\phi$ is too large, 
the effect on $\Delta {N_{\rm eff}^{\rm CMB}}$ of $\phi$ diminishes. 

$\bullet$ { For a particular $m_\phi$ with a constant $y$, we are able to impose a lower bound on the DM mass from the Planck $2\sigma$ 
data on $\Delta N_{\rm eff}^{\rm CMB}$. As an example we find $m_S\gtrsim 7$ MeV when $m_\phi=5$MeV and $y=2\times 10^{-10}$. 
Additionally, a lower bound on $\lambda$ can be derived such that the DM relic abundance remain below 0.12 as experimentally favored.}

In summary, an MeV scale secluded non-minimal dark sector 
is difficult to probe at dedicated dark matter search experiments. In view of this, we propose that the measurement of 
$\Delta N_{\rm eff}^{\rm CMB}$ can be used as a tool to test such a decoupled dark sector. Our work indeed shows that some 
of the parameter space yield observable $\Delta N_{\rm eff}^{\rm CMB}$
and  are already within the reach of the Planck sensitivity. Upcoming CMB stage IV experiments with improved sensitivity 
should be able to probe/refute the allowed 
region of our model parameter space further. 

\section{Acknowledgements}
SG would like to thank Anirban Biswas for some insightful discussions. SG would
also like to thank the University Grants Commission (UGC), Government of India, for providing financial support
as a senior research fellow. AKS is supported by NPDF grant PDF/2020/000797
from Science and Engineering Research Board (SERB), Government of India.

\appendix

\section{Two examples of $SU(2)_L\otimes U(1)_Y$ gauge invariant model}\label{sec:appUV}
In this appendix, we discuss two $SU(2)_L\otimes U(1)_Y$ invariant frameworks that give rise 
to $\phi \overline{\nu}\nu$ vertex effectively at MeV scale.

\vspace{2mm}
\noindent \textbf{Case I:} We consider a variant of type-I seesaw framework with three gauge singlet right 
handed (RH) neutrinos. We impose $\mathbb{Z}_4$ discrete symmetry to write the desired Lagrangian. 
The charge assignments of the relevant fields under the new $\mathbb{Z}_4$ are presented in table\,\ref{tab:modI}. 
We propose the following Lagrangian for generating the $\phi \overline{\nu} \nu$ vertex.
\begin{align}
\mathcal{L}\supset -\dfrac{c_{\phi 1}}{2}\phi\, \overline{N_{R_1}^C}N_{R_1}-\dfrac{c_{\phi 2}}{2}\phi \,\overline{N_{R_2}^C}N_{R_2}-
Y_{e1}\overline{L}_e\widetilde{\Phi}N_{R_1}-Y_{\mu 1}\overline{L}_\mu\widetilde{\Phi}N_{R_2}-
Y_{\tau 3}\overline{L}_\tau\widetilde{\Phi}N_{R_3}-M_{12}\overline{N_{R_1}^C}N_{R_2}-
\frac{M_{33}}{2}\overline{N_{R_3}^C}{N_{R_3}} +\text{ h.c. },\label{eq:ModelI}
\end{align}
where $c_{\phi 1}$, $c_{\phi 2}$ are dimensionless $\mathcal{O}(1)$ coupling coefficients and $\widetilde{\Phi}=i\sigma_2\Phi^*$.

\begin{table}[h]
\begin{center}
\begin{tabular}{ | c | c |}
\hline
Fields & $\mathbb{Z}_4$\\
\hline
$N_{R_1}$ & $i$\\
\hline
$N_{R_2}$ & $-i$\\
\hline
$N_{R_3}$ & $1$\\
\hline
$\phi$ & $-1$\\
\hline
$\{L_e,L_\mu,L_\tau\}$ & $\{i,-i,1\}$\\
\hline
$\Phi$ & 1\\
\hline
\end{tabular} 
\end{center}
\caption{Charge assignment for the relevant fields as proposed in case I. The dark matter field $S$ transforms trivially under the imposed $\mathbb{Z}_4$.}
\label{tab:modI}
\end{table}

   After electroweak symmetry breaking, the SM neutrino mass matrix is given by,
   \begin{align}
   m_\nu\simeq-m_D M_R^{-1} m_D^T,
   \end{align}
   where
   \begin{align}
   m_D=\frac{v}{\sqrt{2}}\begin{pmatrix}
   Y_{e1} & 0 & 0\\
   0 & Y_{\mu 2} & 0\\
   0 & 0 & Y_{\tau 3}
   \end{pmatrix} \text{~~~and~~~}
   M_R=\begin{pmatrix}
   0 & M_{12} & 0\\
   M_{12} & 0 & 0\\
   0 & 0 & M_{33}
  \end{pmatrix},
   \end{align}
The active-sterile neutrino mixing is $\theta_{\alpha i}$, quantified as $\theta_{\alpha i}\simeq \left(m_D M_R^{-1}\right)_{\alpha i}$ \cite{Kersten:2007vk}. 
Due to this active-sterile mixing, the first two terms in Eq.\,\ref{eq:ModelI} can be translated 
to $\phi\overline\nu\nu$ vertex with the coupling coefficient being proportional to $\sim (\theta_{\mu 1}^2 + \theta_{e2}^2)$. 
This set up is able to provide correct order of active neutrino mass ($\sim 0.1$eV) for 
$M_{33},M_{12}\sim \mathcal{O}(1)$GeV and $Y \sim \mathcal{O}(10^{-7})$ with the desired value of $\phi\overline{\nu}\nu$ coupling 
coefficient $\sim \mathcal{O}(10^{-10})$. A detailed study on neutrino masses and mixing is beyond the scope of the present paper.
To prevent the production of DS particles from tree 
level RH neutrino decay ($N \to \phi \nu$), we have assumed that the reheating temperature ($T_{\rm RH}$) of the 
Universe is less than $1$ GeV which is consistent with the lower 
limit on the $T_{\rm RH}$\,($\gtrsim 4$MeV) from big bang nucleosynthesis \cite{Hannestad:2004px}.

\vspace{2mm}
\noindent \textbf{Case II:} We may also start with the following $SU(2)_L\otimes U(1)_Y$ invariant Lagrangian at dimension six level 
to effectively generate the $\phi\overline{\nu}\nu$ vertex \cite{Kouvaris:2014uoa},
\begin{align}
\mathcal{L}\supset -\frac{c_2}{\Lambda^2}\phi\,\overline{\widetilde{L_l}}\,\Phi \,\widetilde{\Phi}^\dagger\, L_l+\text{~h.c.~}\label{eq:modelII},
\end{align}
where $c_2$ is the dimensionless $\mathcal{O}(1)$ coupling coefficient, $\Lambda$ is the cut off scale and $\tilde{L}_l=i\sigma_2L_l^C$ with $l=\{e,\mu.\tau\}$. 
\begin{table}[h]
\begin{center}
\begin{tabular}{ | c | c |}
\hline
Fields & $\mathbb{Z}_4$\\
\hline
$\phi$ & $-1$\\
\hline
$\{L_e,L_\mu,L_\tau\}$ & $\{i,i,i\}$\\
\hline
$\Phi$ & 1\\
\hline
\end{tabular} 
\end{center}
\caption{Charge assignments for the relevant fields as proposed in case II. The dark matter field $S$ transforms trivially under the imposed $\mathbb{Z}_4$.}
\label{tab:modelII}
\end{table}
In this case also, we consider an additional discrete symmetry $\mathbb{Z}_4$ and tabulate the charge 
assignments of the relevant fields in table\,\ref{tab:modelII}. 
After electroweak symmetry breaking one obtains the vertex $\phi\overline{\nu}\nu$ with 
the coupling coefficient being proportional to $\frac{c_2 v^2}{\Lambda^2}$ where $v=246$GeV.

\newpage
\section{Matrix amplitude squares}\label{sec:A2}
\begin{table}[h]
\begin{center}
\begin{tabular}{|c|c|}
\hline
Process & $|\mathcal{M}|^2 /G_F^2$\\
\hline
$\nu_e \bar{\nu}_e \to e^-e^+$ & $-32 (g_L^2 + g_R^2) s u$\\
\hline
$\nu_{\mu (\tau)} \bar{\nu}_{\mu (\tau)} \to e^-e^+$ & $-32 \left[(g_L-1)^2 + g_R^2\right] s u$\\
\hline
$\nu_e e^- \to \nu_e e^-$ & $32 \left[(g_L^2+g_R^2) s^2 + \{(g_L-1)^2+g_R^2\} u^2\right]$\\
\hline
$\nu_{\mu(\tau)} e^- \to \nu_{\mu(\tau)} e^-$ & $32 \left[(g_L-1)^2+g_R^2)\right] (s^2 + u^2)$\\
\hline
$\nu_e e^+ \to \nu_e e^+$ & $32 \left[(g_L^2+g_R^2) u^2 + \{(g_L-1)^2+g_R^2\} s^2\right]$\\
\hline
$\nu_{\mu(\tau)} e^+ \to \nu_{\mu(\tau)} e^+$ & $32 \left[(g_L-1)^2+g_R^2)\right] (s^2 + u^2)$\\
\hline
$\phi\to\overline{\nu}_i\nu_i$ & $\frac{2 \kappa^2 m_\phi^2}{G_F^2}$\\
\hline
\end{tabular}
\caption{Matrix amplitude squares of the relevant processes for neutrino decoupling. Here $G_F$ is the Fermi's coupling constant,
$s,u$ are the Mandelstam variables, $g_L = \dfrac{1}{2} + \sin^2 \theta_W$, $g_R = \sin^2 \theta_W$, and $\theta_W$ is the Weinberg Angle. 
In deriving these amplitudes,
we have neglected the mass of the electron \cite{Kawasaki:2000en,Escudero:2018mvt}.}
\label{table1}
\end{center}
\end{table}
\section{Collision Terms}
\label{sec:A3}
\noindent $\bullet$ The collision term for the evolution of the energy density of $A$ for a 
generic inelastic process such as $A(p_1) + B(p_2) \to C (p_3) + D (p_4)$ is given by
\begin{align}
\mathcal{C}^A_{\rm inel}
=
\int d \Pi_1 d\Pi_2 d\Pi_3 d \Pi_4 E_1 
(2 \pi)^4 \delta^4(p_1 + p_2 - p_3 - p_4)
|\mathcal{M}^2|_{AB\to CD} \Lambda_{\rm inel}(E_1, E_2, E_3, E_4) \,\,,
\end{align}
where 
\begin{align}
\Lambda_{\rm inel} = f_C(E_3) f_D(E_4) (1 \pm f_A(E_1)) (1 \pm f_B(E_2)) - f_A(E_1) f_B(E_2) (1\pm f_C(E_3))(1\pm f_D(E_4)).
\end{align}
Here the distribution function of the $i^{th}$ species is denoted by $f_i$. $p_1$, $p_2$, $p_3$, $p_4$ are four momenta of
$A$, $B$, $C$, $D$ respectively. $|\mathcal{M}^2|_{AB\to CD}$ is the matrix amplitude square of $AB\to CD$ process and 
$d \Pi_i = \dfrac{d^3 \vec{p}_i}{2E_i (2\pi)^3}$.

\vspace{1mm}
\noindent $\bullet$ The collision term for the evolution of the energy density of $A$ for a generic elastic 
process such as $A(p_1) + B(p_2) \to A (p_3) + B (p_4)$ is given by
\begin{align}
\mathcal{C}^A_{\rm el}
=
\dfrac{1}{2}\int d \Pi_1 d\Pi_2 d\Pi_3 d \Pi_4 (E_1  -E_3)
(2 \pi)^4 \delta^4(p_1 + p_2 - p_3 - p_4)
|\mathcal{M}^2|_{AB\to AB} \Lambda_{\rm el}(E_1, E_2, E_3, E_4) ,
\end{align}
where
\begin{align}
\Lambda_{\rm el} = f_A(E_3) f_B(E_4) (1 \pm f_A(E_1)) (1 \pm f_B(E_2)) - f_A(E_1) f_B(E_2) (1\pm f_A(E_3))(1\pm f_B(E_4))\,\,.
\end{align}
Here $|\mathcal{M}^2|_{AB\to AB}$ is the matrix amplitude square of $AB\to AB$ elastic process.

\vspace{1mm}
\noindent $\bullet$ For $A(p_1) \to B (p_2) + C(p_3)$ process, the collision term for the evolution of the energy density of $A$ is as follows:
\begin{align}
\mathcal{C}^A_{A \lra BC} = \int d \Pi_1 d\Pi_2 d\Pi_3 E_1
(2 \pi)^4 \delta^4(p_1 - p_2 - p_3)
|\mathcal{M}^2|_{A\to BC} \Lambda_{A\lra BC}(E_1, E_2, E_3) \,\,,
\end{align}
where $|\mathcal{M}^2|_{A\to BC}$ is the matrix amplitude square for $A \to BC$ process and 
\begin{align}
\Lambda_{A \lra BC} = f_B(E_2) f_C(E_3) (1 \pm f_A(E_1)) - f_A(E_1) (1\pm f_B(E_2))(1\pm f_C(E_3))\,\,.
\end{align}

\bibliography{ref}

\end{document}